\documentclass[a4paper,10pt]{article}

\newcommand{\bfa}[1]{\mbox{\boldmath${#1}$}}

\newcommand{\bea}{\begin{eqnarray}}
\newcommand{\eea}{\end{eqnarray}}
\newcommand{\bnn}{\begin{eqnarray*}}
\newcommand{\enn}{\end{eqnarray*}}
\newcommand{\be}{\begin{equation}}
\newcommand{\ee}{\end{equation}}

\newcommand{\nn}{\nonumber}

\def\PACS{\par\leavevmode\hbox {\it PACS:\ }}%
\def\MSC{\par\leavevmode\hbox {\it MSC:\ }}%
\def\UK{\par\leavevmode\hbox {\it Keywords:\ }}%

\begin{document}

\title{Relativistic quantum physics\\with hyperbolic numbers}

\author{S. Ulrych\\Wehrenbachhalde 35, CH-8053 Z\"urich, Switzerland}
\date{September 25, 2005}
\maketitle

\begin{abstract}
A representation of the quadratic Dirac equation and the Maxwell equations
in terms of the three-dimensional universal complex Clifford algebra
$\bar{\bfa{C}}_{3,0}$ is given. The investigation considers a subset
of the full algebra, which is isomorphic to the Baylis algebra.
The approach is based on the two Casimir operators of the Poincar\'e group,
the mass operator and the spin operator, which is related to the
Pauli-Lubanski vector. The extension to spherical symmetries is discussed briefly. 
The structural difference to the Baylis algebra appears in the shape of the
hyperbolic unit, which plays an integral part in this formalism.
\end{abstract}

{\scriptsize\PACS{12.20.-m; 11.30.Cp; 03.65.Fd; 03.65.Pm; 02.20.Qs}
\MSC{81V10; 11E88; 22E43; 30G35; 20H25}
\UK{Hyperbolic complex Clifford algebra; Hyperbolic numbers; Spin operator; Quadratic Dirac operator; Relativistic spherical symmetries}}

\section{Introduction}
In the same way as complex numbers are associated
with the Euclidean geometry, the two other
systems of bidimensional hypercomplex numbers
can be associated with geometries of physical relevance.
The parabolic numbers can be associated with the Galileo group 
and the hyperbolic numbers with the Lorentz group of special relativity
\cite{Yag79,Sob95}.

The Hamilton quaternions as well as
the bidimensional systems are included in the more general Clifford
algebras \cite{Hes84,Kel94,Cli73,Cli78}. They can represent the Euclidean geometry because 
their invariant quantity is an algebraic quadratic form as well as
the Euclidean and pseudo-Euclidean invariants.

The hyperbolic numbers offer the possibility to represent
the four-compo\-nent Dirac spinor as
a two-component hyperbolic spinor. Hucks has shown \cite{Huc93} that the Lorentz group
is equivalent to the hyperbolic unitary group and that
the operations of C, P, and T on 
Dirac spinors are closely related to the three types of complex conjugation 
that exist when both hyperbolic and ordinary imaginary 
units are present.

Since the relativistic spin group is representable as an unitary group, the special linear
group, which is normally used to represent the relativistic spin, 
has to be considered as an unitary group as well. Porteous \cite{Por81,Por95} proves
the unitarity of the special linear group
with the help of the double field, which
corresponds to the null basis representation of the hyperbolic numbers.

There are various applications of hyperbolic numbers.
They have been applied by Reany \cite{Rea93} to 
2nd-order linear differential equations. A function theory for hyperbolic numbers
has been presented by Motter and Rosa \cite{Mot98}. Extensions to an n-dimensional
space have been given by several authors \cite{Rea94,Fje98,Fje298,Wum02}, including
an analysis of hyperbolic Fourier transformations \cite{Zhe04}.
The functional calculus of hyperbolic numbers is also covered by the 
more general approach of Superanalysis (see e.g. Khrennikov \cite{Khr99}).
Considerations of a non-relativistic hyperbolic Hilbert space with respect to the Born formula
have been given by Kocik \cite{Koc99}, Khrennikov \cite{Khr03,Khr203},
and Rochon and Tremblay \cite{Tre04}.
Xuegang et al. investigated
the Dirac wave equation, Clifford algebraic spinors, a hyperbolic Hilbert space,
and the hyperbolic spherical harmonics in hyperbolic spherical polar coordinates 
\cite{Xue00,Xue200,Xue01}.
The hypercomplex numbers,
and the geometries generated
by these numbers, have been investigated by Catoni et al. \cite{Cat05}.
Further applications of hyperbolic numbers, including e.g.
their application to general relativity by Kunstatter et al., 
can be found in \cite{Cap41,Mac69,Kun83,Fje86,Hes91,Kwa92,Yu95,Xue99}.

It has been shown by Baylis and Jones \cite{Bay89} that a $\bfa{R}_{3,0}$ Clifford algebra
has enough structure to describe relativity as well as
the more usual $\bfa{R}_{1,3}$ Dirac algebra or the $\bfa{R}_{3,1}$ Majorana algebra.
Baylis represents relativistic space-time points as paravectors and applies 
these paravectors to Electrodynamics \cite{Bay99}.
The approach to relativity in terms of $\bfa{R}_{1,3}$ has been investigated
by Hestenes \cite{Hes84,Hes66,Hes03}, or e.g. by Gull, Lasenby, and Doran \cite{Gul93}.
An overview of the structural differences in the above low-dimensional Clifford algebras
is given by Dimakis \cite{Dim89} in the context of a general spinor
representation within Clifford algebras.

The approach used in this work is congruent to Baylis paravectors.
However, hyperbolic numbers are included in the algebra. This corresponds
actually to a hyperbolic complexification of the Baylis algebra. According to
Porteous \cite{Por95} the resulting algebra is
isomorphic to the three-dimensional universal complex Clifford algebra $\bar{\bfa{C}}_{3,0}$.
The correspondance to the Baylis algebra is given by the restriction to
a subset of the algebra. 
This restriction can be justified in the hyperbolic Hilbert space by the hermiticity of 
the Poincar\'e mass operator \cite{Ulr052}.

This work is an extension of \cite{Ulr051}. It introduces beside the
Poincar\'e mass operator an analogeous operator for the spin, which is
related to the Pauli-Lubanski vector. In addition, the approach is 
extended to spherical symmetries.
The mathematical background of the presented
approach with respect to the classification of Clifford algebras  
is not fully explained in \cite{Ulr052,Ulr051}. This will be
improved in this work based on the general overview given by Porteous \cite{Por81,Por95}. The 
notation of Porteous has been adopted in many cases.

\section{Hyperbolic numbers}
\label{not}
Vector spaces can be defined over the commutative ring of 
hyperbolic numbers $z\in\bfa{H}$
\be
\label{beg}
z=x+iy+jv+ijw\;,\hspace{0.5cm}x,y,v,w \in\bfa{R}\;,
\ee
where the complex unit $i$ and the hyperbolic unit $j$ have the properties
\be
i^2=-1\;,\hspace{0.5cm}j^2=1\;.
\ee
The hyperbolic numbers defined in this way are a commutative extension of the complex numbers to
include new roots of the polynomial equation $z^2 - 1 = 0$.
In the terminology of Clifford algebras
they are represented by $\bar{\bfa{C}}_{1,0}$, i.e.
they correspond to the universal one-dimensional complex Clifford algebra
(the notation follows Porteous \cite{Por95}).

Beside the grade involution, two anti-involutions play a major role in the
description of Clifford algebras and their structure, conjugation and reversion.
Conjugation changes the sign of the complex and
the hyperbolic unit
\be
\label{conj}
\bar{z}=x-iy-jv+ijw\;.
\ee 
The hyperbolic numbers with conjugation
are equivalent to the double
field $^2\bar{\bfa{C}}^\sigma$ of Porteous, where $\sigma$ and the bar symbol denote
swap and complex conjugation. The notation of Porteous has the advantage that it can clearly specify, whether
the double field is defined over the real $^2\bfa{R}$, complex $^2\bfa{C}$, 
or quaternionic numbers $^2\bfa{Q}$. If necessary the notation used in this work
is extended to $\bfa{H}(\bfa{R})$,
$\bfa{H}(\bfa{C})=\bfa{H}$, or $\bfa{H}(\bfa{Q})$, where $\bfa{H}(\bfa{R})$ and 
$\bfa{H}(\bfa{Q})$ correspond to the universal Clifford algebras 
$\bfa{R}_{1,0}$ and $\bfa{R}_{0,3}$, respectively.

With respect to the Clifford conjugation the square of the 
hyperbolic number can be calculated as
\be
\label{square}
z\bar{z}=x^2+y^2-v^2-w^2+2ij(xw-yv)\;,
\ee
i.e. in general the square of a hyperbolic number is not a
real number. 

Beside the conjugation, the second important anti-involution is the reversion, which
changes only the sign of the complex unit
\be
\label{cconj}
z^{\dagger}=x-iy+jv-ijw\;.
\ee
Anti-involutions reverse the ordering in the multiplication,
e.g. $(ab)^\dagger=b^\dagger a^\dagger$. This becomes
important when non-commuting elements of an algebra are considered.
In physics, reversion is denoted as hermitian conjugation. Note, that in \cite{Ulr052} it has been
suggested to relate hermiticity in the physical sense to the conjugation anti-involution.

\section{Hyperbolic algebra}
\label{vec}
Consider a hyperbolic vector with the coordinates $z^\mu=(z^0,z^i)\in\bar{\bfa{H}}^{\,3,1}$.
The bar symbol indicates that the vector has the signature $(3,1)$ in the hermitian
product with respect to conjugation, i.e. $z_\mu \bar{z}^\mu=z_0\bar{z}^0-z_i\bar{z}^i$.
The vector can be represented in terms of a Clifford algebra as
\be
\label{veco}
Z=z^\mu e_\mu\;.
\ee
The basis elements $e_\mu=(e_0,e_i)$ include the
unity and the Pauli algebra
multiplied by the hyperbolic
unit~$j$ 
\be
 e_\mu=(1,j\sigma_i)\;.
\ee
The pseudoscalar of the hyperbolic algebra, which will appear thoughout this work,
is defined as
\be
I=e_0\bar{e}_1e_2\bar{e}_3=ij\;.
\ee

Important for the further analysis is the behaviour of the above hypercomplex units under 
conjugation and reversion. In Table \ref{invo} it is displayed whether the sign
of the unit is changed or not under the considered operation.
For completeness the graduation as an example of an important involution is displayed. 
An involution
does not change the ordering in a product, i.e. $\widehat{ab}=\hat{a}\hat{b}$.
The graduation is used to identify the even elements of a Clifford algebra.
A certain subset of these elements defines the spin group of the considered Clifford algebra \cite{Por95}.

\begin{table}
\begin{center}
\begin{tabular}{|c|c|c|c|}
\hline
$a$ & $\bar{a}$ & $a^\dagger$ & $\hat{a}$ \\
\hline
$e_i$ & $-$ & $+$ & $-$\\
\hline
$\sigma_i$ & $+$ & $+$ & $+$\\
\hline
$I$ & $+$ & $-$ & $-$\\
\hline
$i$ & $-$ & $-$ & $+$\\
\hline
$j$ & $-$ & $+$ & $-$\\
\hline
\end{tabular}
\end{center}
\caption{Effect of conjugation, reversion, and graduation on the used hypercomplex units.\label{invo}}
\end{table}

Adding the hyperbolic unit to the Pauli algebra corresponds in fact to a
hyperbolic complexification.
In the terminology of Clifford algebras the complexification is leading to the
isomorphisms
\be
\bfa{R}_{3,0}\otimes\bar{\bfa{H}}(\bfa{R})
\simeq\bar{\bfa{H}}(\bfa{R})_{3,0}\simeq\bfa{H}(2)\simeq\bar{\bfa{C}}_{3,0}\;.
\ee
The full structure therefore corresponds to the universal three-dimensional complex Clifford
algebra. The representation of the Clifford algebra in terms of matrices is given by
$\bfa{H}(2)$, the algebra of hyperbolic $2\times 2$ matrices. This representation will
be assumed in some of the equations within this work.

The vector $Z$ of Eq.~(\ref{veco}) has sixteen real dimensions.
In \cite{Ulr052} it has been shown that the four-dimensional
real Minkowski vector can be considered as the magnitude of $Z$,
if the square is restriced to real numbers, i.e. $Z\bar{Z}\in\bfa{R}$.
Based on the assumption that only such vectors are of physical relevance,
the following
investigation is restricted to the real four-dimensional Minkowski space,
and the remaining phase contributions are neglected.
A Minkowski vector $x^\mu\in\bfa{R}^{\,3,1}$ is expressed in the above algebra
as $X=x^\mu e_\mu$.

The spatial vector contributions can be written explicitly in the
following representation
\bea
X&=&x^0+j\bfa{x}\;,
\eea
where $\bfa{x}=x^i\sigma_i$.
Using the Pauli matrices as the explicit representation of $\sigma_i$, 
the vector $X$ can be expressed in terms of a hyperbolic
$2\times 2$ matrix according to 
\be
X=\left(\begin{array}{cc}
\;x^0+jx^3\;&\;jx^1-ijx^2\;\\
\;jx^1+ijx^2\;&\;x^0-jx^3\;\\
\end{array}\right)\;.
\ee
The scalar product of two vectors
can be defined as
\be
\label{scalar}
X\cdot Y
=\frac{1}{2}(X\bar{Y}+Y\bar{X})\;.
\ee 
The wedge product is given as
\be
\label{outer}
X\wedge Y
=\frac{1}{2}(X\bar{Y}-Y\bar{X})\;.
\ee
The wedge product corresponds to a so-called biparavector, which can be
used e.g. for the description of the electromagnetic field or
the relativistic angular momentum (see also Baylis \cite{Bay99}).

The basis elements of the relativistic $\bar{\bfa{C}}_{3,0}$ paravector algebra can be considered
as the basis vectors of the relativistic vector space. 
These basis elements form a non-cartesian orthogonal basis with respect to the
scalar product defined in Eq.~(\ref{scalar}) 
\be
 e_{\mu}\cdot e_{\nu}
=g_{\mu\nu}\;,
\ee
where $g_{\mu\nu}$ is the metric tensor of the Minkowski space.

As an example of the paravector algebra
the energy-momentum vector of a free 
classical pointlike particle, moving 
with velocity 
$\bfa{v}$ relative to the observer, 
is expressed in the Pauli algebra notation.
The relativistic momentum vector for this 
particle can be written as 
\be
\label{koko}
P=\frac{E}{c}+j\bfa{p}=mc\exp{(j\bfa{\xi})}\;,
\ee
with $c$ denoting the velocity of light,
$\bfa{\xi}$ the rapidity, $E$ the energy and
$\bfa{p}$ the momentum of the particle. 
The rapidity is defined as $tanh\xi=v/c=pc/E$,
where $\xi=|\bfa{\xi}|$ and $p=|\bfa{p}|$.
Rapidity and momentum 
point into the same direction $\bfa{n}=\bfa{v}/|\bfa{v}|$ as the velocity.
In the following $c$ and $\hbar$ will be set equal to one.
In quantum mechanics the momentum 
is replaced in coordinate space by the operators $p^\mu=i\partial^\mu$.

\section{Lorentz transformations}
\label{trafo}
Porteous \cite{Por95} shows that the general linear group can be
considered as an unitary group. The unitarity is related to the double field 
together with an appropriate correlation. For the complex linear group this is a $^2\bar{\bfa{C}}^\sigma$-correlation. 
The complex double field $^2\bar{\bfa{C}}^\sigma$, with swap $\sigma$ 
and conjugation, correponds to the null basis representation of the 
hyperbolic numbers with conjugation $\bar{\bfa{H}}$ as given
in Eqs.~(\ref{beg}) and (\ref{conj}).
In addition, there is an isomorphism between the general linear group and the hyperbolic
unitary group $GL(n,\bfa{C})\simeq U(n,\bfa{H})$, 
and in particular for the special groups in two dimensions, i.e. $SL(2,\bfa{C})\simeq SU(2,\bfa{H})$
(see also Hucks \cite{Huc93}).
The unitarity of the group $U(n,\bfa{H})$ is understood with respect to a $\bar{\bfa{H}}$-correlation.

The group $SU(2,\bfa{H})$ corresponds to the spin group of
$SO(3,1,\bfa{R})$ and its elements can
be used to express rotations and boosts of the
paravectors defined in the last section.
The rotation of a paravector can be expressed as \cite{Bay99}
\be
\label{rota}
X\rightarrow X^\prime=R X\, R^\dagger\;.
\ee
For the boosts one finds the transformation rule
\be
\label{boost}
X\rightarrow X^\prime= B X B^\dagger\;.
\ee
The rotations and boosts are given as
\be
\label{rotmat}
R=\exp{(-i\bfa{\theta}/2)}\;,\hspace{0.5cm}B=\exp{(j\bfa{\xi}/2)}\;.
\ee

Based on the Pauli matrices an explicit matrix representation
of the boost operator $B$ can be given, e.g. for a boost in the direction of the
$x$-axis one finds
\be
B_1=\left(\begin{array}{cc}
\;\cosh{\xi_1/2}\;&\;j\sinh{\xi_1/2}\;\\
\;j\sinh{\xi_1/2}\;&\;\cosh{\xi_1/2}\;\\
\end{array}\right)\;.
\ee
The boosts are invariant under reversion
$B^\dagger = B$, whereas the conjugated
boost corresponds to the inverse
$\bar{B}=B^{-1}$. For rotations reversion and conjugation correspond both to the inverse
$R^\dagger=\bar{R}=R^{-1}$. This relationship indicates that in non-relativistic physics
hermitian operators can be defined either
with respect to reversion or conjugation.

Boosts and rotations
can be combined to form the Lorentz transformation
\be
\label{lorentz}
X\rightarrow\; X^\prime= L X L^\dagger\;,
\ee
which can be expressed in terms of its
infinitesimal generators as
\be
\label{lorentzmat}
L=\exp{\left(-i\theta^iJ_i-i\xi^iK_i\right)}\;.
\ee
From these equations the infinitesimal generators of a Lorentz
transformation can be identified as
\be
\label{gener}
\bfa{J}=\bfa{\sigma}/2\;,\hspace{0.5cm}\bfa{K}= ij\bfa{\sigma}/2\;. 
\ee
One can show 
that the generators satisfy the  
Lie algebra of the Lorentz
group $SO(3,1,\bfa{R})$.
The Lorentz transformations can be expressed also with relativistic second
rank tensors
\be
\label{relspin}
L=\exp{(-iS_{\mu\nu}\omega^{\mu\nu}/2)}\;,
\ee
where the spin angular momentum is defined in terms of the wedge product as
\be
\label{spinwedge}
e_\mu\wedge e_\nu=2S_{\mu\nu}\;.
\ee
The elements of the spin angular momentum therefore represent
planes in space-time that are formed by the basis elements of the
algebra.

The following example shows how coordinate vector, momentum vector, orbital angular momentum
and spin angular momentum can be related to each other in the paravector algebra
\be
\label{ospinref}
X\bar{P}=x_\mu p^\mu -iS_{\mu\nu}L^{\mu\nu}\;,
\ee
where $L^{\mu\nu}=x^\mu p^\nu-x^\nu p^\mu$ corresponds to
the relativistic orbital angular momentum.

\section{Poincar\'e mass operator}
\label{secmassop}
With the above vector representation the Poincar\'e mass operator can be introduced as a product of a momentum vector and
its conjugated counterpart
\be
\label{pmassop}
M^2=P\bar{P}\;.
\ee

The explicit form of the mass operator is obtained by a multiplication of the basis matrices.
The mass operator can be separated into a spin dependent and a
spin independent contribution  
\be
\label{spindef}
P\bar{P}=p_\mu p^\mu-i\sigma_{\mu\nu}p^\mu p^\nu\;,
\ee
where the spin term is given by
\be
\label{sigspin}
\sigma_{\mu\nu}=
\left(\begin{array}{cccc}
\;0\;&\;-ij\sigma_1\;&\;-ij\sigma_2\;&\;-ij\sigma_3\;\\
\;ij\sigma_1\;&\;0\;&\;\sigma_3\;&\;-\sigma_2\;\\
\;ij\sigma_2\;&\;-\sigma_3\;&\;0\;&\;\sigma_1\;\\
\;ij\sigma_3\;&\;\sigma_2\;&\;-\sigma_1\;&\;0\;\\
\end{array}\right)\;.
\ee
Since the spin contribution is anti-symmetric, the last term in Eq.~(\ref{spindef}) 
is in this case zero. The spin structure becomes important
when interactions are introduced by the minimal substitution of the momentum operators.
The anti-symmetric contribution $\sigma_{\mu\nu}$ correponds to the wedge product of Eq.~(\ref{spinwedge}),
which gives the analogous relation $\sigma_{\mu\nu}=2S_{\mu\nu}$.

The basic fermion equation is introduced as an eigenvalue equation of the mass operator.
With the hyperbolic algebra defined above the
equation can be written as
\be
\label{equat}
M^2\psi(x)=m^2\psi(x)\;,
\ee
The wave function $\psi(x)$ has the general structure 
\be
\label{Ansatz}
\psi(x)=\varphi(x)+j\chi(x)\;,
\ee
where $\varphi(x)$ and $\chi(x)$ can be represented as two-component
spinor functions (see Hucks \cite{Huc93}). They depend on the four space-time
coordinates $x^\mu$.

\section{Poincar\'e spin operator}
\label{planewave}
In analogy to the mass operator
a spin operator can be introduced. 
The spin operator
corresponds to the second Casimir operator of the Poincar\'e group, which
describes a system that is invariant under relativistic
translations and rotations. The basic equation for the spin operator 
can be defined as
\be
\label{spinequat}
S^2\psi(x)=-s(s+1)\psi(x)\;,
\ee
where the square of the spin operator corresponds to 
\be
\label{pspinop}
S^2=W\bar{W}/m^2\;.
\ee
The operator $W$ denotes the Pauli-Lubanski vector, which
can be expressed in terms of the complex Clifford algebra as
\be
\label{Luba3}
W=-\bfa{J}\cdot\bfa{p}-j\,(\bfa{J}p^0+\bfa{K}\times\bfa{p})\;.
\ee
The spatial spin operator can be derived
from the Pauli-Lubanski vector by projections (see Michel \cite{Mic59} or Wightman \cite{Wig60})
\be
\bfa{S}=\frac{1}{m}\,W\cdot\bfa{n}\;.
\ee
The projection vectors $n^{\mu}$ are an arbitrary set of four orthogonal vectors 
satisfying the relation $n^{\mu}\cdot n^{\nu}=g^{\mu\nu}$.
The spin operator takes the following form
\be
\label{spinoperator}
\bfa{S}=\frac{1}{m}\left(\bfa{J}p^0+\bfa{K}\times\bfa{p}-(\bfa{J}\cdot\bfa{p})
\frac{\bfa{p}}{p^0+m}\right)\;,
\ee
if the set of projection vectors is chosen as
\bea
n^0&=&m^{\!-1}(p^0,p^k)\;,\nonumber\\
n^i&=&m^{\!-1}(p^i,m\delta^{ik}+p^ip^k/(p^0+m))\;.
\eea
The eigenstates of the spin operator
can be introduced as eigenvectors of the squared spin operator and of the
z-component $S_z=S_3$
\bea
S^2\;\vert\,s\,m_s\,\rangle
&=&-s(s+1)\;\vert\,s\,m_s\,\rangle\;,
\\
S_z\;\vert\,s\,m_s\,\rangle
&=&m_s\;\vert\,s\,m_s\,\rangle
\nonumber\;.
\eea

To find an explicit representation of these eigenstates one has to consider
that the spin operator in the above form
can be obtained also with a boost acting
on the operator vector $\bfa{J}$
\be
\label{shref}
\bfa{S}=B \bfa{J}\bar{B}\;.
\ee
Therefore, the relativistic spinor can be 
related to the non-relativistic
Pauli spinor by a boost
\be
\label{begin}
\vert\,s\,m_s\,\rangle
=B\chi_{m_s}=u(\bfa{p},m_s)\;.
\ee 
If the rapidity in
the boost (see Eq.~(\ref{rotmat}))
is expressed in terms of the particle momentum, one finds that
the boosted Pauli spinor corresponds to the
hyperbolic representation of the Dirac spinor
\be
\label{spnra}
u(\bfa{p},m_s)=\sqrt{\frac{p^0+m}{2m}}\left(1
+\frac{j\bfa{p}}
{p^0+m}\right)\chi_{m_s}\;.
\ee
The anti-particle spinor can be derived from the
particle spinor, which is multiplied by the 
hyperbolic unit $v(\bfa{p},m_s)=ju(\bfa{p},m_s)$.
The spinors can be combined with the Hilbert space state for the momentum
$\vert p^\mu\rangle$ to form the plane wave
expansion
\begin{eqnarray}
\label{solute}
\psi(x)&=&\sum_{m_s}\int \! \frac{d^3\bfa{p}}{(2\pi)^32p^0}\,\left(
u(\bfa{p},m_s)e^{-ip_\mu x^\mu}\,b(p,m_s)\right.\nonumber\\
&&+\left.v(\bfa{p},m_s)e^{ip_\mu x^\mu}\,\bar{d}(p,m_s)\right)\;,
\end{eqnarray}
which is a general solution of the Poincar\'e mass operator and
the Poincar\'e spin operator~(\ref{spinequat}).

The relativistic wave function is an element of the spinor space,
which is a minimal left ideal in the terminology of Clifford algebras. 
The elements of a minimal left ideal
have rank~1 and therefore they can be represented as column vectors
$\psi^i(x)\in \bar{\bfa{H}}^2$. The transformation rule of the spin operator in Eq.~(\ref{shref}) is in
contrast to Eq.~(\ref{boost}). The above rule is used for relativistic operators acting in the
spinor space. The transformation rule is related to the $\bar{\bfa{H}}$-correlation, which maps
the elements of the spinor space to their dual space. 
Since the Poincar\'e mass operator and
the Poincar\'e spin operator are operators in the spinor space they have to obey the same transformation law. 
This is the reason why e.g. the mass
operator is defined as $P\bar{P}$ and not as $PP^\dagger$.

\section{Massless particles}
\label{massless}
Massless particles like neutrinos are normally described in terms of helicity states.
It is shown that massless particles can also be described with
boosted Pauli spinors. 
Representing particles in terms of Pauli spinors corresponds to
a decoupling of the polarization axis and the direction of momentum.
In explicit calculations this decoupling is leading to 
ambiguities in the coupling of angular momenta of multiple particles. Therefore, e.g. in
calculations of scattering amplitudes, the
helicity basis is the best choice, even in a non-relativistic scheme. 

From the theoretical point of view, 
massless particles that are represented by Pauli spinors provide a direct 
analogy to the description of the last
section. What has to be shown is that this representation is conform with the
Poincar\'e group in the massless case
\be
\label{equat5}
M^2\psi(x)=0\;.
\ee
If the polarization axis for a massive particle is chosen, e.g. as the z-axis, the momentum 
can take any direction without any restriction. This is not the case for massless
particles. The spatial momentum can never be perpendicular to the chosen
polarization axis. The spatial symmetry of the momentum is
therefore broken, which will become apparent in the following equations.

The spinor will be defined within the little
group of a standard vector \cite{Tun85}. Since massless particles are moving
with the velocity of light they have no rest frame. Therefore, 
the standard frame will be defined as the system in which
the momentum is directed along the polarization
axis. If the particle is polarized in the direction of the z-axis
the positive-energy standard vector is given as 
\be
\label{standml}
p^\mu=(\vert p^3\vert,0,0, p^3)\;.
\ee 
In this description the helicity can be either positive or negative. The states are
always characterized by the two possible polarizations corresponding to the
chosen polarization axis. The components of the Pauli-Lubanski
vector in the standard frame are
\bea
\label{PLu}
W^0&=&-p^3 J^3\nonumber\;,\\
W^1&=&-\vert p^3\vert(J^1+K^2)\;,\\
W^2&=&-\vert p^3\vert(J^2-K^1)\nonumber\;,\\
W^3&=&-\vert p^3\vert J^3\nonumber\;.
\eea 
The operators $W^0$ and $W^3$ are linear dependent and can
be represented by $J^3$. 
The three generators $J^3$, $W^1$ and $W^2$ satisfy the
Lie algebra of $E_2$, the Euklidean group in two dimensions, 
which defines the little group of 
the $m^2=0$ representation.

An arbitrary momentum vector $p^\mu$ is obtained with
a boost perpendicular to the z-axis
\be
\label{rapidy}
p^\mu=(\vert p^3\vert \cosh{\xi}\,,\,\vert p^3\vert\, n^1 \sinh{\xi}\,,\,
\vert p^3\vert\, n^2\sinh{\xi}\,,p^3)\;.
\ee
The unit vector $n^i_\perp=(n^1,n^2,0)$ 
characterizes the direction of the boost. 
With the perpendicular momentum vector $p_\perp^i=(p^1,p^2,0)$
the rapidity can be defined by the relation $\tanh{\xi}=p_\perp/p^0$,
where $p_\perp=\vert \bfa{p}_\perp\vert$.
Using $\bfa{\xi}_\perp=\bfa{n}_\perp\xi$ the boost can 
be written in the form $B=\exp{(j\bfa{\xi}_\perp/2)}$.
The momentum contribution parallel
to the polarization will be denoted as $p_\parallel^i=(0,0,p^3)$, with $p_\parallel=\vert \bfa{p}_\parallel\vert$.

In the spinor representation the generators $\bfa{J}=\bfa{\sigma}/2$ and $\bfa{K}=ij\bfa{\sigma}/2$
can be inserted into Eq.~(\ref{PLu}).
Then one finds in the standard frame and therefore in all
frames
$W\bar{W}=0$,
i.e.~the spin is given in the
degenerate spin $s=0$ representation of $E_2$.
The basis vectors are chosen as eigenvectors 
of $J_3$ with the eigenvalues $m_s=\pm \frac{1}{2}$. They can be represented
with the Pauli spinor $\chi_{m_s}$. A general spinor
can be calculated using Eq.~(\ref{begin}) with the 
boost parameters defined above
\be
\label{m0spnra}
u(\bfa{p},m_s)=\sqrt{\,\frac{p^0+p_\parallel}{2p_\parallel}\,}\left(1
+\frac{j\bfa{p}_\perp}
{p^0+  p_\parallel }\right)\chi_{m_s}\;.
\ee
The antiparticle spinor is obtained if the
above expression is multiplied
by the hyperbolic unit. 

A set of four
orthogonal projection vectors can be introduced to derive the spin operator
from the Pauli-Lubanski vector
\bea
n^0&=&  p_\parallel^{-1}(p^0,p^k_\perp)\;,\nonumber\\
n^i&=&  p_\parallel^{-1}(p_{\perp}^i,  p_\parallel \delta^{ik}+p_{\perp}^i p^k_\perp/(p^0+  p_\parallel ))\;.
\eea
The spin operator is then defined as
\be
\label{nspin}
\bfa{S}=\frac{1}{  p_\parallel }\, W\cdot \bfa{n}\;.
\ee
The spin operator corresponds again to a boosted vector of the
angular momentum generators $\bfa{J}$. Explicitly written one finds
\be
\bfa{S}= \frac{1}{  p_\parallel }
\left(\bfa{J}p^0+\bfa{K}\times\bfa{p}_\perp-(\bfa{J}\cdot\bfa{p}_\perp)\frac{\bfa{p}_\perp}{p^0+p_\parallel}\right)\;.
\ee 
As mentioned above only the third component $S_z=S_3$ of the spin operator is relevant within $E_2$.
For this component the last term in the above equation is zero.
The action of the spin operators on the spinor can be summarized in the
relations
\bea
\label{ndef}
S^2\;\vert\,0\,m_s\,\rangle
&=&0\;,
\\
S_z\;\vert\,0\,m_s\,\rangle
&=&m_s\;\vert\,0\,m_s\,\rangle
\nonumber\;,
\eea
where the square of the spin operator corresponds to $S^2=W\bar{W}/p_\parallel^2 $.
The plane wave expansion is formally equivalent to
Eq.~(\ref{solute}), but 
the spinors $u(\bfa{p},m_s)$ and $v(\bfa{p},m_s)$
have to be replaced 
with the specific $m^2=0$ form
given above. From these equations it follows that
\be
S^2 \psi(x)=0\;.
\ee
In fact, massless particles are also spinless particles. However, they can be
described by a non-zero momentum and a polarization comparable to the
massive particles.
 
\section{Maxwell equations}
\label{electro}
The Maxwell equations can be derived from an eigenvalue equation of the mass operator,
where the mass operator is now acting on a vector field
\be
\label{wavb}
M^2 A(x)=0\;.
\ee
The equation can be expressed
with the electromagnetic fields according to
\bea
\label{maxwell}
P\bar{P}A&=&-\bfa{\nabla}\cdot\bfa{E}-\partial^0 C\nonumber\\
                  &&+ij\bfa{\nabla}\cdot\bfa{B}\\
                  &&-j(\bfa{\nabla}\times\bfa{B}-
                       \partial^0\bfa{E}
                      -\bfa{\nabla}C)\nonumber\\
                  &&-i(\bfa{\nabla}\times\bfa{E}+\partial^0\bfa{B})
                   =0\nn\;.
\eea
This expression is obtained if one evaluates $\bar{P}A(x)$, inserts the usual definitions for the
electromagnetic fields, and then multiplies the resulting terms by the operator
$P$. If $P\bar{P}$ is calculated first, Eq.~(\ref{wavb}) reduces to the wave operator acting on the
vector potential giving zero. Both forms are equivalent in the Lorentz gauge. 

In Eq.~(\ref{maxwell}) the four homogeneous 
Maxwell equations are included. 
The calculation provides two additional terms depending on 
\be
\label{zero}
C(x)=\partial_\mu\, A^\mu(x)\;.
\ee
These terms disappear in the Lorentz gauge. 

\section{Photon plane wave states}
\label{photop}
In this section a plane wave expansion for free
photon fields is derived. The techniques
developed for massless fermions will be applied. In this representation
the polarization
vector of the photon corresponds to a generalization of the Pauli spinor. Again it is mentioned that the
helicity basis has advantages in explicit calculations.

The transformation properties of the vector 
components $A^\mu(x)$ can be understood in terms of 
$4\times 4$ transformation
matrices acting on four-component vectors 
\bea
\label{lorentz2}
x^\mu\rightarrow x^{\mu\prime}= (L)^{\mu}_{\;\;\nu}\,x^\nu\;.
\eea
The transformation matrices can be derived from the generators
$J_i$ and $K_i$ according to Eq.~(\ref{lorentzmat}).
The third components of the generators are
\be
\label{photongen}
(J_3)^{\mu}_{\;\;\nu}=\left(\begin{array}{cccc}
\;0\;&\;0\;&\;0\;&\;0\;\\
\;0\;&\;0\;&\;-i\;&\;0\;\\
\;0\;&\;i\;&\;0\;&\;0\;\\
\;0\;&\;0\;&\;0\;&\;0\;\\
\end{array}\right),
\hspace{0.5cm} 
(K_3)^{\mu}_{\;\;\nu}=\left(\begin{array}{cccc}
\;0\;&\;0\;&\;0\;&\;i\;\\
\;0\;&\;0\;&\;0\;&\;0\;\\
\;0\;&\;0\;&\;0\;&\;0\;\\
\;i\;&\;0\;&\;0\;&\;0\;\\
\end{array}\right).
\ee
The results of Section \ref{massless} will be used in the following.
The standard frame is the system in which the momentum
is directed along the polarization axis.
The Pauli-Lubanski vector in the standard frame
is given by Eq.~(\ref{PLu}), where the generators now have to
be replaced by the generators of Eq.~(\ref{photongen}), including the
remaining generator components. 

The two eigenstates of 
$J_3$ with eigenvalues $m_s=\pm 1$ are given as usual as
\be
\chi_{\pm}^\mu
=(\mp\, \epsilon_1^\mu - i\epsilon_2^\mu)/\sqrt{2},
\ee
where $\epsilon_1^\mu$ and $\epsilon_2^\mu$ are
unit vectors in the direction of the x- and y-axis. 
In the standard frame one finds that the squared spin operator acting on
the eigenstates is zero
\be
(W \bar{W})
^{\mu}_{\;\;\nu}\,\chi_{m_s}^\nu=0.
\ee
General eigenstates of the spin operator are obtained again by a boost of the
eigenstate from the standard frame of the particle $\vert\,0\,m_s\,\rangle=e^\mu(\bfa{p},m_s)=(B)^{\mu}_{\;\;\nu}\,\chi_{m_s}^\nu$.
As in section \ref{massless} the direction of the boost is perpendicular
to the polarization axis, i.e. $B=\exp{(-i\xi^i_\perp K_i/2)}$. The boost is
performed with the generators
given in Eq.~(\ref{photongen}). The calculation is leading to the 
explicit form of the polarization vector
\be
e^\mu(\bfa{p},m_s)=\left(\,\frac{p^{m_s}}{  p_\parallel }\,,
\chi_{m_s}^i+ \frac{p^i_\perp p^{m_s}}{  p_\parallel (p^0+  p_\parallel )}
\right)\;,
\ee
where $p^\pm=(\mp p^1-
ip^2)/\sqrt{2}$. 
For the eigenstates one finds the relations
\bea
S^2\;\vert\,0\,m_s\,\rangle
&=&0\;,
\\
S_z\;\vert\,0\,m_s\,\rangle
&=&m_s\;\vert\,0\,m_s\,\rangle
\nonumber\;.
\eea

The spin operator $S_z=S_3$
is defined as in Eq.~(\ref{nspin}) with the appropriate
generators for $\bfa{J}$ and $\bfa{K}$. The above equations
are equivalent to Eq.~(\ref{ndef}) except  
for the different eigenvalues. For photons
one finds $m_s=\pm 1$. 
The plane wave expansion of the
free photon field is given as  
\be
\label{solutex}
A(x)=
\sum_{m_s}\int \! \frac{d^3\bfa{p}}{(2\pi)^32p^0}\,
e^\mu (\bfa{p},m_s)\,e_\mu\!\left(e^{-ip_\mu x^\mu}a(p,m_s)+
e^{ip_\mu x^\mu}\,\bar{a}(p,m_s)\right).
\ee
The plane wave satisfies the relation
\be
S^2 A(x)=0\;.
\ee

The coordinate vector $\chi_{m_s}^\mu\in \bar{\bfa{C}}^{\,3,1}$ plays in this context the role
of the Pauli spinor. It can be considered as an element of the minimal left ideal with
respect to the transformations induced by the generators of Eq.~(\ref{photongen}). 
This is indicated in the above formulas by the tensor indices.
The indices can be omitted, e.g. $e(\bfa{p},m_s)=B\chi_{m_s}$,
to provide a representation free
notation.
 
\section{Quadratic Dirac equation}
\label{blab1}
Electromagnetic interactions can be introduced with the minimal substitution of the
momentum operator. The mass operator of Eq.~(\ref{pmassop}) transforms into
\be
\label{basic}
M^2=(P-eA(x))(\bar{P}-e\bar{A}(x))\;,
\ee
and is now invariant under local gauge transformations.
This mass operator can be inserted into Eq.~(\ref{equat}). If Pauli matrices 
and electromagnetic fields are expressed with the
anti-symmetric tensor $\sigma_{\mu\nu}$ given in Eq.~(\ref{sigspin})
and $F^{\mu\nu}=\partial^\mu A^\nu- \partial^\nu A^\mu$ one finds
\be
\label{Pauli1}
\left((p-eA)_\mu(p-eA)^\mu -\frac{e}{2}
\sigma_{\mu\nu} F^{\mu\nu}-m^2\right)\psi(x)=\,0\;.
\ee
This is the quadratic Dirac equation. Using the hyperbolic algebra
it can be represented as a $2\times 2$ matrix equation, whereas conventionally
the quadratic Dirac equation is a $4\times 4$ matrix equation.
The spinor function $\psi(x)$
used for the mass operator has a two-component structure, 
whereas in the Dirac equation $\psi(x)$ corresponds to a four-component spinor.

One finds in both cases the same two coupled
differential equations. 
For the hyperbolic algebra one can derive from Eq.~(\ref{Pauli1})
\be
\label{Pauli}
\left((p-eA)_\mu(p-eA)^\mu -e\,ijE^i\sigma_i
+eB^i\sigma_i -m^2\right)\psi(x)=0\;.
\ee
In the Dirac theory the spin tensor is defined 
according to $\sigma_{\mu\nu}=i/2\,[\gamma_\mu,\gamma_\nu]$.
Using this tensor the Dirac form of Eq.~(\ref{Pauli}) can be written as
\be
\label{Pauli3}
\left((p-eA)_\mu(p-eA)^\mu -
e\,iE^i\alpha_i
+eB^i\sigma_i -m^2\right)\psi(x)=0\;.
\ee
Comparing this equation with Eq.~(\ref{Pauli}) one observes
that in both cases the term including the electric field is the only
term which couples the components
of the spinor. In the mass operator
equation the coupling term
is proportional to $j\bfa{\sigma}$, in the quadratic Dirac equation
the term corresponds to 
$\bfa{\alpha}=\gamma_5\bfa{\sigma}$.
The Dirac
representation of $\gamma_5$ and $j$ have
the same effect on the spinor, a swap of the spinor components.
For the Poincar\'e mass operator one finds
\be
j\psi(x)=\chi(x)+j\varphi(x)\;,
\ee
whereas in the Dirac representation the corrsponding relation is given as
\be
\gamma_5\psi(x)=
\left(\begin{array}{c}
\;\chi(x)\;\\ \;\varphi(x)
\end{array}\right)\;.
\ee

In the hyperbolic formalism
the terms proportional to the hyperbolic unit include the
differential equation of the lower component, the other terms describe the
differential equation of the upper component. Conventionally,
the coupled differential equations for upper and lower components are separated by
the matrix structure.

\section{Orbital angular momentum and single particle potentials}
\label{orbitan}
The hyperbolic numbers
can be used also for the description of the
orbital angular momentum, which will be shown in this section
(compare with Xuegang \cite{Xue00}).
A spacelike relativistic vector $x^\mu$ can be parametrized
in relativistic spherical coordinates as
\be
\label{para}
x^{\mu}=\left(\begin{array}{c}
x^{0}\\
x^{1}\\
x^{2}\\
x^{3}\\
\end{array}\right)=
\left(\begin{array}{c}
\rho\, \sinh{\xi}\\
\rho\, \cosh{\xi}\,\sin{\theta}\,\cos{\phi}\\
\rho\, \cosh{\xi}\,\sin{\theta}\,\sin{\phi}\\
\rho\, \cosh{\xi}\,\cos{\theta}\\
\end{array}\right)\;.
\ee
The time coordinate is given as
$\rho\, \sinh{\xi}$, where $\rho > 0$. In the limit of $\xi\rightarrow 0$ the
vector reduces to a non-relativistic vector in spherical
coordinates. 

Based on the Lorentz transformation given
in Eqs.~(\ref{lorentz2}) and (\ref{photongen}) the above vector can be obtained 
from a standard vector $x^\mu=(0,0,0,\rho)$ with the 
following transformation
\be
\label{lform}
L(\theta,\phi,\xi)= \exp{(-iJ_3\,\phi\,)}\,
\exp{(-iJ_2\,\theta\,)}\,\exp{(-iK_3\,\xi\,)}\;. 
\ee

For an 
irreducible group representation of the orbital angular momentum
the relation between the generators of
boosts and rotations
is assumed to be the same as for the spin $(\frac{1}{2},0)$ representation 
of the Lorentz group used earlier in this work
\be
\bfa{K}=ij\bfa{J}.
\ee
The boost generator in Eq.~(\ref{lform})
can then be replaced by the third component of the rotation generator 
\be
\label{lform2}
L(\theta,\phi,\xi)= \exp{(-iJ_3\,\phi\,)}\,
\exp{(-iJ_2\,\theta\,)}\,\exp{(jJ_3\,\xi\,)}\;.
\ee
The orbital angular momentum will be denoted by $\bfa{L}$ in contrast to the
spin angular momentum $\bfa{S}$. In the following equations $\bfa{J}$ will be specialized to
$\bfa{J}=\bfa{L}$. 
The $(l,0)$ representation of the Lorentz group provides the following
equations for the irreducible states
\bea
L^2\;\vert\,l\,m_l\,\rangle
&=&-l(l+1)\;\vert\,l\,m_l\,\rangle\;,
\\
L_z\;\vert\,l\,m_l\,\rangle
&=&m_l\;\vert\,l\,m_l\,\rangle
\nonumber\;.
\eea
Again the third component is denoted as $L_z=L_3$.
In this basis the transformation of Eq.~(\ref{lform}) is represented as
\bea
D^l_{m_l^\prime m_l}(\theta,\phi,\xi)&=&
\langle\,l\,m_l^\prime\,\vert\,e^{-iL_3\,\phi\,}\,
e^{-iL_2\,\theta\,}\,e^{\,jL_3\,\xi\,}\,\vert\,l\,m_l\,\rangle\nonumber\\
&=&e^{-im_l^\prime\,\phi\,}\,d^{\,l}_{m_l^\prime m_l}
(\theta\,)\,e^{\,jm_l\xi\,}\;.
\eea
Compared to the non-relativis\-tic
case the relativistic rotation matrices 
are extended by the additional hyperbolic phase factor
$e^{\,jm_l\xi\,}$.

An application of these functions can be the solution of the Poincar\'e mass operator
with appropriate model potentials.
A relativistic generalization of the $1/ \vert \bfa{x}\vert$ 
central potential is suggested, which could be used to
describe an electron moving in the potential of a nucleus
\be
\label{newpot}
eA^\mu(x)= -\frac{Z\alpha}{\rho}\,\epsilon^\mu(x)\;.
\ee
The polarization vector corrsponds to the unit vector
of the $\xi$-coordinate
\be
\epsilon^\mu(x)= \frac{1}{\rho}\,\frac{\partial}
{\partial\,\xi}\,x^\mu(\rho,\theta,\phi,\xi)
=
\left(\begin{array}{c}
\cosh{\xi}\\
\sinh{\xi}\,\sin{\theta}\,\cos{\phi}\\
\sinh{\xi}\,\sin{\theta}\,\sin{\phi}\\
\sinh{\xi}\,\cos{\theta}\\
\end{array}\right)
\;.
\ee
In the static limit $\xi\rightarrow 0$ this potential reduces
to the $1/ \vert \bfa{x} \vert$ potential.

Some general remarks on the solution
procedure will be given here.
The $\bfa{\sigma}$-terms
in Eq.~(\ref{Pauli}) 
imply a coupling of the orbital angular momentum with the spin
$\bfa{J}=\bfa{L}+\bfa{S}$. 
There is also a term proportional to the hyperbolic unit 
which has a
different parity. The wave function $\psi(x)$ therefore has a similar
structure as the Dirac spinor
\be
\psi(x)=\varphi_l(x)+ j\chi_{l^\prime}(x)\;,
\ee
with
\be
l^\prime=
\left\{\begin{array}{c}
l-1\\
l+1\\
\end{array}\right.
\hspace{0.5cm}\mathrm{for}\hspace{0.5cm}
\begin{array}{c}
l=j+1/2\\
l=j-1/2\\
\end{array}\;.
\ee
The eigenvalue of the coupled spin operator should not be confused with
the hyperbolic unit in this equation.
Since relativistic single particle potentials depend on the relative time,
the energy of the single particle states is not a 
conserved quantity. Therefore, the following eigenvalue problem has to be considered
\be
M^2\;\psi(x)= m^2\;\psi(x)\;,
\ee
where the solutions
$\psi_i(x)$
are eigenstates of the mass operator
with the  quantum numbers $i=(njlm_j)$.
Given the model potential of
Eq.~(\ref{newpot}) the eigenvalues $m_i^2$ of this equation should be close to the
spectrum of the Dirac equation with a non-relativistic $1/ \vert \bfa{x}\vert$ 
central potential. For the ground state $GS=(1\frac{1}{2}0m_j)$
one can therefore expect to find approximately
\be
m_{GS}^2\approx 1-Z^2\alpha^2\;.
\ee 
A detailed solution of this problem is beyond the scope of this
work.

\section{Summary}
The representation of relativistic quantum physics in terms of mathematical
structures is not unique. The relationships and isomorphisms between different
representations can be understood within Clifford algebras as the common 
underlying mathematical framework.
The most popular representation of relativistic physics is based on the Dirac algebra $\bfa{R}_{1,3}$.
However, the Majorana algebra $\bfa{R}_{3,1}$, or the $\bfa{R}_{3,0}$ algebra suggested by Baylis,
provide frameworks that can be used for relativistic calculations as well.

The three-dimensional universal complex Clifford algebra $\bar{\bfa{C}}_{3,0}$ is
proposed for an alternative representation of relativistic quantum physics.
The Baylis algebra is isomorphic to the subset of
$\bar{\bfa{C}}_{3,0}$ considered in this work. The structural difference
appears in the shape of the hyperbolic unit. The full structure of the
complex $\bar{\bfa{C}}_{3,0}$ Clifford algebra provides sixteen real dimensions, the same number as
the Dirac algebra, which is in contrast to
the eight-dimensional Baylis algebra.

\appendix
\section{Relationship to other representations}
The algebra introduced in Eq.~(\ref{veco}) seems to be identical to the
$\bfa{R}_{3,0}$ paravector algebra of Baylis \cite{Bay89,Bay99}. Especially, if
one considers the corresponding anti-involutions in Table \ref{invo} for the basis elements $e_i$. However, Baylis
identifies the basis elements $e_i$ directly with the Pauli algebra. In his approach the
Pauli algebra therefore has different transformation properties under the
anti-involutions than in the approach presented here. 

The hyperbolic algebra can be represented also in terms of quaternions
as 
\be
\label{quat}
 e_\mu=(1,ij \,q_i)\;, 
\ee
where $q_i\in\bfa{Q}$ denote the basis elements of the
quaternion algebra. The three-dimensional vector symbol $\bfa{x}$ has been represented as
$x^i \sigma_i$, because the Pauli algebra is the one most familiar to physicists. 
It could be represented also as $x^i e_i$ or as
$x^i q_i$. 
It is possible to change from one picture to the other
if $j\sigma_i$ is replaced by $e_i$ or $ijq_i$
including a redefinition of $\bfa{x}$.


\begin{thebibliography}{0}
\bibitem{Yag79}
I. M. Yaglom, {\it A Simple Non-Euclidean Geometry and its Physical Basis},
(Springer Verlag, New York, 1979).
\bibitem{Sob95}
G. Sobczyk, {\it The College Mathematics Journal}, September (1995).
\bibitem{Hes84}
D. Hestenes, G. Sobczyk, {\it Clifford Algebra to geometric calculus}, (Reidel, Dordrecht, 1984).
\bibitem{Kel94}
J. Keller, {\it Advances in Applied Clifford Algebras} {\bf 4} (1), 1 (1994).
\bibitem{Cli73}
W. K. Clifford, {\it Proc. Lond. Math. Soc.} {\bf 4}, 381 (1873).
\bibitem{Cli78}
W. K. Clifford, {\it Am. J. Math.} {\bf 1}, 350 (1878).
\bibitem{Huc93}
J. Hucks, {\it J. Math. Phys.} {\bf 34}, 5986 (1993).
\bibitem{Por81}
I. Porteous, {\it Topological Geometry}, 1st ed., (Van Nostrand Reinhold, London, 1969).
\bibitem{Por95}
I. Porteous, {\it Clifford algebras and the classical groups}, (Cambridge University Press, Cambridge, 1995).
\bibitem{Rea93} 
P. Reany, {\it Advances in Applied Clifford Algebras} {\bf 3} (2), 121 (1993). 
\bibitem{Mot98}
A. E. Motter, M. A. F. Rosa, {\it Advances in Applied Clifford Algebras} {\bf 8} (1), 109 (1998).
\bibitem{Rea94}
P. Reany, {\it Advances in Applied Clifford Algebras} {\bf 4} (1), 89 (1994).
\bibitem{Fje98}
P. Fjelstad, S. Gal, {\it Advances in Applied Clifford Algebras} {\bf 8} (1), 47 (1998).
\bibitem{Fje298}
P. Fjelstad, S. Gal, {\it Advances in Applied Clifford Algebras} {\bf 8} (2), 309 (1998).
\bibitem{Wum02}
L. Wuming, {\it Advances in Applied Clifford Algebras} {\bf 12} (1), 7 (2002).
\bibitem{Zhe04}
Z. Zheng, Y. Xuegang, , {\it Advances in Applied Clifford Algebras} {\bf 14} (1), 207 (2004).
\bibitem{Khr99}
A. Khrennikov, {\it Superanalysis},
(Kluwer Academic Publishers, Dordrecht, 1999).
\bibitem{Koc99}
J. Kocik,
{\it International Journal of Theoretical Physics} {\bf 38} (8), 2221 (1999).
\bibitem{Khr03}
A. Khrennikov, {\it Advances in Applied Clifford Algebras} {\bf 13} (1), 1 (2003).
\bibitem{Khr203}
A. Khrennikov, {\it Annalen der Physik} {\bf 12} (10), 575 (2003).
\bibitem{Tre04}
D. Rochon, S. Tremblay, {\it Advances in Applied Clifford Algebras} {\bf 14} (1), 231 (2004).
\bibitem{Xue00}
Y. Xuegang, {\it Advances in Applied Clifford Algebras} {\bf 10} (1), 49 (2000).
\bibitem{Xue200}
Y. Xuegang, L. Wuming, {\it Advances in Applied Clifford Algebras} {\bf 10} (2), 163 (2000).
\bibitem{Xue01}
Y. Xuegang, Z. Shuna, H. Qiunan, {\it Advances in Applied Clifford Algebras} {\bf 11} (1), 27 (2001).
\bibitem{Cat05}
F. Catoni, R.Cannata, V. Catoni, P. Zampetti, {\it Advances in Applied Clifford Algebras} {\bf 15} (1), 1 (2005).
\bibitem{Cap41} 
P. Capelli,
{\it Bull. of American Mathematical Society} {\bf 47}, 585 (1941).
\bibitem{Mac69}
G. Mack, A. Salam, {\it Annals of Physics} {\bf 53}, 174 (1969).
\bibitem{Kun83}
G. Kunstatter, J. W. Moffat, J. Malzan, {\it J. Math. Phys.} {\bf 24}, 886 (1983).
\bibitem{Fje86}
P. Fjelstad, {\it American Journal of Physics} {\bf 54} (5), 416 (1986).
\bibitem{Hes91}
D. Hestenes, P. Reany, G. Sobczyk, {\it Advances in Applied Clifford Algebras}
{\bf 1} (1), 51 (1991).
\bibitem{Kwa92}
A. K. Kwa\~sniewski, {\it Advances in Applied Clifford Algebras} {\bf 2} (1), 107 (1992).
\bibitem{Yu95}
Y. Xuegang, Y. Xuequian, {\it Acta Mathematica Scientia} {\bf 15} (4), 435 (1995).
\bibitem{Xue99}
Y. Xuegang, {\it Advances in Applied Clifford Algebras} {\bf 9} (1), 109 (1999).
\bibitem{Bay89}
W. E. Baylis,  G. Jones, {\it Journal of Physics A} {\bf 22}, 1 (1989).
\bibitem{Bay99}
W. E. Baylis, {\it Electrodynamics: A Modern Geometrical Approach}, (Birk\-hauser, Boston, 1999).
\bibitem{Hes66}
D. Hestenes, {\it Space Time Algebra},
(Gordon and Breach, New York, 1966).
\bibitem{Hes03}
D. Hestenes, American Journal of Physics {\bf 71}, 691 (2003).
\bibitem{Gul93}
S. Gull, A. Lasenby, C. Doran, {\it Found. Phys.} {\bf 23} (9), 1175 (1993).
\bibitem{Dim89}
A. Dimakis, {\it Journal of Physics A} {\bf 22} 3171 (1989).
\bibitem{Ulr052}
S. Ulrych, {\it Physics Letters B} {\bf 618}, 233 (2005).
\bibitem{Ulr051}
S. Ulrych, {\it Physics Letters B} {\bf 612}, 89 (2005).
\bibitem{Mic59}
L. Michel, {\it Il Nouvo Cimento Supplemento} {\bf 14}, 95 (1959).
\bibitem{Wig60}
A. S. Wightman in 
{\it Relations de Dispersions et Particules \'El\'ementaires},
eds. C. DeWitt and M. Jacob
(Hermann and John Wiley, New York, 1960). 
\bibitem{Tun85}
Wu-Ki Tung, {\it Group Theory in Physics},
(World Scientific, Singapore, 1985).
\end{thebibliography}
\end{document}